\documentclass[aps,onecolumn,groupedaddress,notitlepage]{revtex4-1}

\def\d{{\rm d}}\def\r{{\bf r}}

\def\e{{\bf e}}
\def\f{{\bf f}}
\def\p{{\partial}}

\usepackage{graphicx}  % needed for figures
\usepackage{dcolumn}   % needed for some tables
\usepackage{bm}        % for math
\usepackage{amssymb,amsfonts,amsmath}   % for math
\usepackage{color}
\usepackage{lipsum,subeqnarray}

\usepackage[normalem]{ulem}

\renewcommand{\d}{\mathrm{d}}

\renewcommand{\d}{\mathrm{d}}

\allowdisplaybreaks

\begin{document}

\title{Elastohydrodynamic Synchronization of Adjacent Beating Flagella}

\author{Raymond~E.~Goldstein}
\author{Eric~Lauga}
\author{Adriana~I.~Pesci}
\author{Michael R.E.~Proctor}
\affiliation{Department of Applied Mathematics and Theoretical Physics, Centre for Mathematical Sciences, \\ University of 
Cambridge, Wilberforce Road, Cambridge CB3 0WA, United Kingdom}

\date{\today}

\begin{abstract}  It is now well established that nearby beating pairs of eukaryotic flagella or cilia typically synchronize in phase.
A substantial body of evidence supports the hypothesis that hydrodynamic coupling between the active filaments, combined with waveform compliance, provides 
a robust mechanism for synchrony.  This elastohydrodynamic mechanism has been incorporated into `bead-spring' models in which the beating 
flagella are represented by microspheres tethered by radial springs as they are driven about orbits by internal forces.  While these low-dimensional models
reproduce the phenomenon of synchrony, their parameters are not readily relatable to those of the filaments they represent.  More realistic models which
reflect the underlying elasticity of the axonemes and the active force generation, take the form of fourth-order nonlinear PDEs.  While computational 
studies have shown the occurrence of synchrony, the effects of
hydrodynamic coupling between nearby filaments governed by such continuum models have been theoretically 
examined only in the regime of interflagellar distances $d$ large
compared to flagellar length $L$.  Yet, in many biological situations $d/L \ll 1$.  Here, we first present an asymptotic analysis 
of the hydrodynamic coupling between two extended filaments in the regime $d/L \ll 1$, and find that the form of the coupling is independent of the 
microscopic details of the internal forces that govern the motion of the individual filaments.  The analysis is analogous to that yielding the 
localized induction approximation for vortex filament motion, extended to the case of mutual induction.  In order to understand how the 
elastohydrodynamic coupling mechanism leads to synchrony of extended objects, we introduce a heuristic model of flagellar beating.  The model
takes the form of a single fourth-order nonlinear PDE whose form is derived from symmetry considerations, the physics of elasticity, and the
overdamped nature of the dynamics.  Analytical and numerical studies of this model illustrate how synchrony between a pair of filaments 
is achieved through the 
asymptotic coupling.

\end{abstract}

\maketitle

\section{introduction}
\label{sec:intro}
\begin{figure}[t]
\centering
\includegraphics[width=0.6\textwidth]{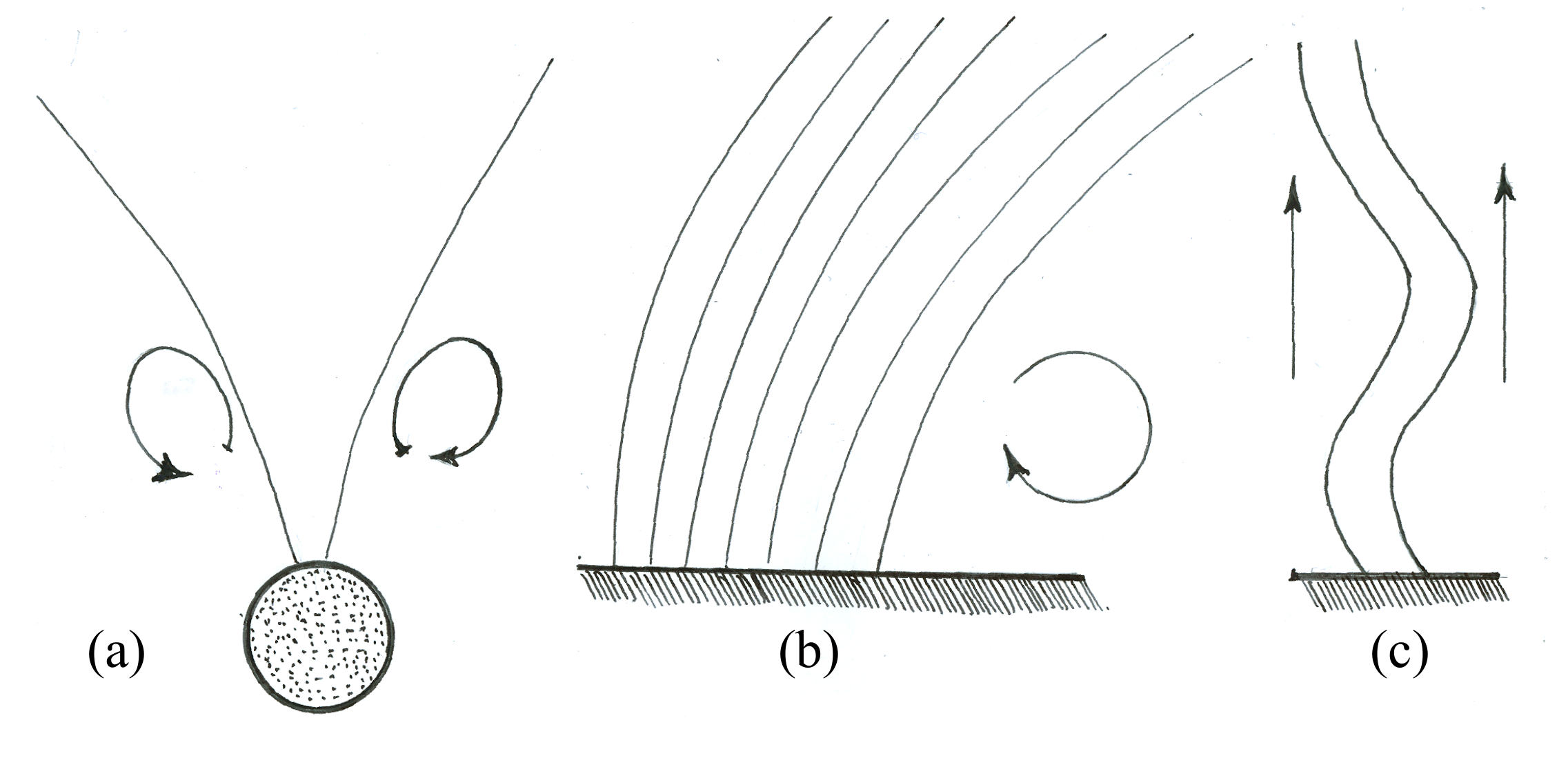}
\caption{A variety of flagellar beating patterns. (a) breaststroke of the biflagellate {\it Chlamydomonas}, (b) ciliary beating in a 
metachronal wave, (c) photoshock response of {\it Chlamydomonas}.}
\label{fig1}
\end{figure}

In nearly all of the contexts in biology in which groups of cilia or flagella are found they exhibit some form of synchronized behavior. 
At the level of unicellular organisms this often takes the form of precise phase synchrony as in the breaststroke beating of biflagellated 
green algae \cite{ARFM}, but it has also been known since the work of Rothschild \cite{Rothschild} that swimming sperm cells can 
synchronize the beating
of their tails when they are in close proximity.  In multicellular organisms such as \textit{Paramecium} \cite{Brennen} 
and \textit{Volvox} \cite{Volvox_metachronal1,Volvox_metachronal2}, and in the respiratory and reproductive systems of higher animals 
one often observes 
metachronal waves, which are long-wavelength modulations in the beating of ciliary carpets.  There are three primary
dynamical behaviors of ciliary groups, two involving beating waveforms that have 
a `power stroke' in which the filament pivots as
a nearly straight rod, followed by a `recovery stroke' in which it is strongly curved, and a third in which the flagella beating is
undulatory.  In the first two cases it is useful to categorize the different
geometries on the basis of the orientation of the power strokes of adjacent flagella.  If we follow reference points along each
flagellum - say, each center of mass - then they will move with either parallel (cilia) or anti-parallel (biflagellate) angular velocities 
(Figs. \ref{fig1}a \& b).  
In the undulatory case (Fig. \ref{fig1}c), found in mutants of \textit{Chlamydomonas} and during the `photoshock response' 
\cite{photoshock}, nearby flagella beat parallel to each other.  

Based on the observations of Rothschild \cite{Rothschild} on synchronized swimming of nearby sperm cells, Taylor \cite{Taylor} 
investigated the possibility that hydrodynamic 
interactions could lead to synchrony. His `waving sheet model' considered two 
infinite parallel sheets each in the shape of a prescribed unidirectional sinusoidal traveling wave.
Examining the rate of viscous dissipation as a function of the phase shift between the two waves, he found that the synchronized state
had the least dissipation.  While highly plausible as an explanation of synchronization, this model does not include a dynamical
mechanism by which the synchronized state is achieved from arbitrary initial conditions.  Subsequent work \cite{Elfring} has shown
that adding waveform flexibility to the model yields a true dynamical evolution toward synchrony, and this has been
confirmed by experiment \cite{eLife}. The recognition that hydrodynamic interactions
alone are insufficient to generate dynamical evolution toward synchrony, and that some form of generalized flexibility is necessary had 
already been seen in the study of rotating helices as a model for bacterial flagella \cite{Powers}.
This notion of `orbital compliance' was subsequently incorporated into several variants of bead-spring models
\cite{NEL,Vilfan,BroutCicuta}
of ciliary dynamics in which each beating filament is replaced by a moving microsphere which is driven around an orbit by internal forces
and allowed to deviate by a radial spring.  Under the assumption that radial motions evolve rapidly relative to azimuthal ones, these
models generically yield a nonlinear ODE for the phase difference between the oscillators that takes the form of the Adler
equation \cite{Adler}.

While these models lead to a microscopic interpretation of the generic Adler equation, they are most appropriate for the situation in which the
distance $d$ between the flagella is large compared to their length $L$, where a far-field description in terms of Stokeslets is valid \cite{Guirao}.  
But many of the most interesting situations, such as the parallel and undulating geometries in Figs. \ref{fig1}b\&c, are in precisely the 
opposite limit, $d/L\ll 1$, while still in the regime $a\ll d$, where $a$ is the filament radius.  In this limit it is clear 
that a proper description of the entire filament is necessary, 
because there are very strong
near-field interactions all along them, and therefore a representation by a single point force would not be realistic.  
While computations incorporating microscopic models of flagella embedded in a viscous fluid show that synchronization does indeed occur through
hydrodynamic interactions in this regime \cite{Yang,Mitran}, it was only in subsequent work that 
the formally exact nonlocal description of hydrodynamic interactions in multi-filament systems was presented \cite{TornbergShelley}.
Taking advantage of the separation of scales $a\ll d\ll L$, we present in Section II an asymptotic 
derivation of the leading-order hydrodynamic coupling between two filaments.  In particular, we find that the relevant small coupling
parameter is $\epsilon=\frac{\ln(L/d)}{\ln(L/a)}$.    The analysis leading to this result
is reminiscent of the `localized induction approximation' in vortex filament dynamics \cite{LIA}.

At present, there is no single generally-accepted microscopic model for eukaryotic flagellar beating, although recent studies have begun 
to address the relative merits of several promising candidate models \cite{BaylyWilson_bj,BaylyWilson_int,Howard}, which typically
consist of a pair of coupled equations, one for the filament displacement, incorporating filament bending elasticity and viscous drag,
and the other for the active bending forces associated with molecular motors.  To illustrate how the hydrodynamic coupling 
and waveform compliance lead to synchronization, in Section III we introduce and analyze a heuristic single PDE,
of the form $(h_i)_t ={\cal N}(h_i); \  i= 1,2$, where ${\cal N}$ is a nonlinear operator, which displays
self-sustained finite-amplitude wavelike solutions.  Section IV considers, both analytically and numerically, 
the dynamics of two filaments of the type introduced in Section III interacting through the coupling derived in Section II.  The concluding
Section V discusses possible applications of the model.

\section{asymptotics}
\label{sec:asym}

\begin{figure}[t]
\centering
\includegraphics[width=0.70\textwidth]{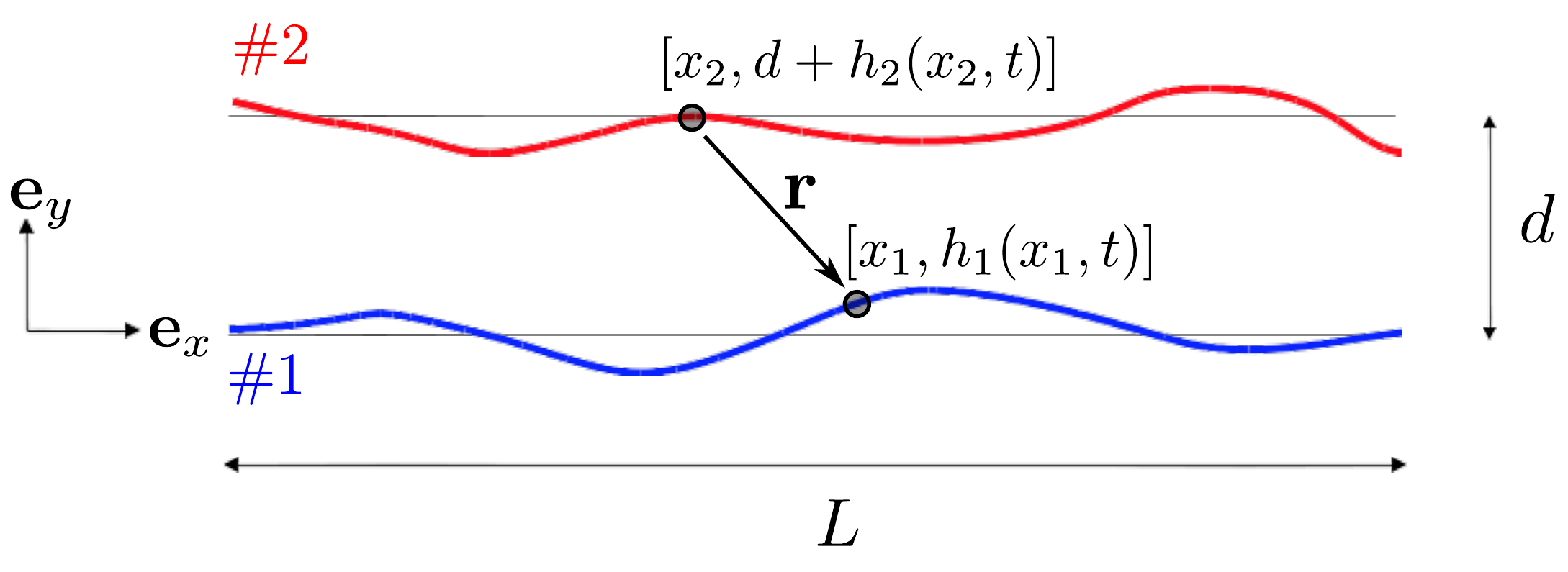}
\caption{Notation for the calculation.  Two filaments, each of length $L$, are separated by a mean distance $d$.  The 
displacements from the mean of two arbitrary points on the curves, separated by the 
vector ${\bf r}$, are $h_1(x_1,t)$ and $h_2(x_2,t)$.}
\label{fig2}
\end{figure}

We consider two slender filaments of length $L$ undergoing some waving motion, as shown in Fig. \ref{fig2}. Their mean separation is 
$d$ and we assume that their waving amplitude is at most of order $d$. We seek to derive, in the linear regime of small amplitude 
displacements, the forces resulting from hydrodynamic 
interactions in the asymptotic limit $d\ll L$. 

Let $h_1$ and $h_2$ denote the vertical displacements 
of the filaments from their mean positions.  We present the derivation of the force on one filament only,
say filament 1, as the dynamics of the other can be deduced by symmetry. In the linear regime, it is only necessary 
to consider the balance of forces in the vertical direction. Using the framework of resistive-force theory (RFT) \cite{Cox}, 
the vertical component $F_1$ of the hydrodynamic force per unit length acting at the point $(x_1, h_1)$ of filament $1$ is 
\begin{equation}
F_1=-\zeta_\perp \left[\frac{\p h_1}{\p t}(x_1,t) - u_y^{2\to 1}(x_1,h_1) \right],
\end{equation}
where $\zeta_\perp$ is the drag coefficient for motion normal to the filament.  
We now proceed to calculate the flow $u_y^{2\to 1}(x_1,h_1)$ induced by filament 2. 
In the linear regime, this flow arises from a superposition of hydrodynamic point forces acting in the $y$ direction 
along filament 2. Neglecting end effects, the flow $u_y^{2\to 1}$ can thus be written as the following integral of Stokeslets:
\begin{equation}
u_y^{2\to 1}(x_1,h_1)=\e_y\cdot \int_0^L
\frac{1}{8\pi \mu}\left(\frac{\bf 1}{r }+\frac{\r\r}{r^3}\right)\cdot \f_2
\,\d x_2~,
\end{equation}
where $\mu$ is the dynamic viscosity, ${\bf 1}$ the identity, and $\r$ the vector that points from $(x_2, d+h_2)$ on filament 2 towards $(x_1, h_1)$ on 
filament 1 (Fig.~\ref{fig1}).  In the linear regime, the force density is simply given 
by that from RFT as
\begin{equation}
\f_2 = \zeta_\perp \frac{\p h_2}{\p t}(x_2,t)\e_y,
\end{equation}
where the positive sign indicates that the force $\f_2$ acts on the fluid. We thus have
\begin{equation}
u_y^{2\to 1}(x_1,h_1)=\frac{\zeta_\perp}{8\pi \mu}  \int_0^L
\left[\frac{1}{r }+\frac{(\r\cdot \e_y)^2}{r^3}\right]\frac{\p h_2}{\p t} 
\,\d x_2.
\label{4}
\end{equation}
Substituting $\r=[x_1-x_2,h_1-d-h_2]$ into Eq.~\eqref{4} and linearizing for small $h_1$ and $h_2$, we obtain
\begin{equation}\label{6}
u_y^{2\to 1}(x_1,h_1)\simeq \frac{\zeta_\perp}{8\pi \mu}  \int_0^L
\frac{(x_2-x_1)^2+2d ^2}{[(x_2-x_1)^2 
+ d ^2]^{3/2}}\frac{\p h_2}{\p t} 
\,\d x_2 + \dots.
\end{equation}

In order to compute the asymptotic value of Eq.~\eqref{6} in the limit $d\ll L$ we introduce 
the dimensionless lengths $x_{i}'=x_i/L,~ i=1,2$, define $\epsilon_1=d/L$, and obtain
\begin{equation}
I(x_1)= \int_0^1
\frac{(x_2-x_1)^2+2\epsilon_1^2}{[(x_2-x_1)^2 
+ \epsilon_1^2]^{3/2}}\frac{\p h_2}{\p t} 
\,\d x_2,
\label{pelotas}
\end{equation}
where for convenience we have dropped the prime on the integration variable, while all other quantities in (\ref{pelotas}) are
still dimensional. This integral has two contributions: a local integral, $I_L$, 
where $x_2$ is close to $x_1$ and a nonlocal 
one, $I_{NL}$ where $|x_2-x_1|\gg \epsilon_1$.   To evaluate $I_{NL}$, we choose an intermediate
length scale $\delta$ satisfying $\epsilon_1 \ll \delta \ll 1$.   Then, 
the nonlocal integral can be rewritten as
\begin{equation}
I_{NL}(x_1)= \left(\int_0^{x_1-\delta} + \int_{x_1+\delta}^{1}\right)
\frac{(x_2-x_1)^2+2\epsilon_1^2}{[(x_2-x_1)^2 
+ \epsilon_1^2]^{3/2}}\frac{\p h_2}{\p t} 
\,\d x_2.
\end{equation}
By construction $|x_1-x_2|$ is at least $\delta\gg \epsilon_1$ and thus it is possible 
to neglect $\epsilon_1$ in the nonlocal integral to obtain
\begin{equation}\label{9}
I_{NL}(x_1)\simeq  \int_0^{x_1-\delta} 
\frac{1}{(x_1-x_2)
}\frac{\p h_2}{\p t} 
\,\d x_2
+ \int_{x_1+\delta}^{1}
\frac{1}{(x_2-x_1 )
}\frac{\p h_2}{\p t} 
\,\d x_2~.
\end{equation}
Provided that $x_1(1-x_1) \gg \delta^2$, then in the limit $\delta \to 0$, the integral in Eq.~\eqref{9} diverges 
logarithmically as
\begin{equation}\label{10}
I_{NL}(x_1)
= - \left[2 \ln \delta + O(1)\right] \frac{\p h_2}{\p t} \bigg|_{x_2=x_1}.
\end{equation}

Similarly, and with the change of variables $\Delta = x_2-x_1$, the local part of the integral can be written as
\begin{equation}
I_{L}(x_1)= \int_{-\delta}^{\delta}
\frac{\Delta^2+2\epsilon_1^2}{[\Delta^2 
+ \epsilon_1^2]^{3/2}}\frac{\p h_2}{\p t} 
\,\d \Delta,
\end{equation}
where $h_2$ is now understood to be a function of $\Delta$. 
In the limit $\delta\to 0$, the term ${\p h_2}/{\p t} $ takes its value near $\Delta=0$, and
the remaining integral can be computed exactly, yielding
\begin{equation}
I_{L}(x_1)\simeq \left\{
\ln\left[\frac{\sqrt{\delta^2 + \epsilon_1^2} + \delta}{\sqrt{\delta^2 + \epsilon_1^2} - \delta}
\right] + 
\frac{2\delta }{[\delta^2 
+ \epsilon_1^2]^{1/2}}\right\}
\frac{\p h_2}{\p t} \bigg|_{\Delta=0}.
\end{equation}
Because $\epsilon_1 \ll \delta$ the result to order $\epsilon_1 $ reads \begin{equation}\label{15}
I_{L}(x_1)\simeq \left[
2\ln \delta + 2 \ln 2-2\ln \epsilon_1
+ 2 + O\left(\frac{\epsilon_1^2}{\delta^2}\right)
\right]
\frac{\p h_2}{\p t} \bigg|_{\Delta=0}.
\end{equation}

Remarkably, the $\ln \delta$ divergence from Eq.~\eqref{10} exactly cancels out the one from Eq.~\eqref{15} producing 
a result which is independent of the particular choice made for the cutoff $\delta$. Thus,
the final expression for $I$ reads
\begin{equation}
I(x_1) = [- 2 \ln \epsilon_1+O(1)]\frac{\p h_2}{\p t} \bigg|_{x_2=x_1},
\end{equation}
Thus, 
the vertical component of the hydrodynamic force on filament $1$ is
\begin{equation}\label{18}
F_1\simeq -\zeta_\perp \left\{\frac{\p h_1}{\p t}\bigg|_{x_1,t} 
- \frac{\zeta_\perp}{4\pi \mu} \ln\left( \frac{L}{d}\right) 
\frac{\p h_2}{\p t} \bigg|_{x_2=x_1,t} \right\}\simeq 
-\zeta_\perp \left\{\frac{\p h_1}{\p t}\bigg|_{x_1,t} 
- \epsilon
\frac{\p h_2}{\p t} \bigg|_{x_2=x_1,t} \right\},
\end{equation}
where we have used $\zeta_{\perp} \sim 4\pi \mu/\ln (L/a)$, with $a$ the radius of the flagella, and 
\begin{equation}
\epsilon\equiv\frac{\ln(L/d)}{\ln(L/a)}~.
\end{equation}
As in nearly all applications of slender body hydrodynamics, the parameter $\epsilon$ is asymptotically small only in
the unphysical case when the aspect ratio is exponentially large.  However, it is well-known that use of the leading order
approximation of slender body hydrodynamics for larger values of $\epsilon$ is robust \cite{Batchelor,PowersLauga}. 

Finally, it is important to note that even in the case when the 
filaments are in phase and $h_1(x_1,t)=h_2(x_2,t)=\tilde h(x,t)$ everywhere, and where
\begin{equation}\label{21}
F_1\simeq -\zeta_\perp \left(1-\epsilon\right)
  \frac{\p \tilde h}{\p t}\bigg|_{x,t},
\end{equation}
it is not appropriate to 
evaluate Eq.~\eqref{21} at close contact between the two flagellar filaments ($d=2a$).
because the induced flow was computed as a superposition of Stokeslets only.  This is a good
approximation only when all other singularities present have decayed away, in particular the (potential) $1/r^3$ 
source dipole which arises 
from the finite-size of the flagella.  Thus, the approximation requires that $d \gg a$ from every point of flagella 1 to every 
point in flagella 2 (and vice-versa). In other words, the result \eqref{18} is valid only within the limits $a \ll d \ll L$.  
For example, for eukaryotic flagella with $L\sim 50$ $\mu$m and $a\sim 0.1$ $\mu$m, 
then the analysis is valid when the flagella are separated by a few microns, in which case $\epsilon$ decreases from $0.5$ for 
$d\sim 2$~$\mu$m to $0.25$ for $d\sim 10$~$\mu$m.

The results above imply that if each of two nearby filaments is governed by an equation of the form 
$\partial h_i/\partial t ={\cal N}_{c_i}(h_i)$, where $\{c_i\}$ are the parameters that differentiate the flagella, then
the coupled pair evolves according to
\begin{subeqnarray}
\label{coupled0}
{\partial h_1 \over \partial t} &=& {\cal N}_{c_1}(h_1) + \epsilon  {\partial h_2 \over \partial t}, \\
{\partial h_2 \over \partial t} &=& {\cal N}_{c_2}(h_2) + \epsilon  {\partial h_1 \over \partial t}. \ \ 
\end{subeqnarray}
As we have only computed the hydrodynamic interaction to order $\epsilon$, it is appropriate to consider the leading-order
form of \eqref{coupled0} as $\partial h_i/\partial t={\cal N}_{c_i}(h_i) + \epsilon  {\cal N}_{c_j}(h_j)$ for $i,j=1,2$ and $i\ne j$.

\section{phenomenological model of a single filament}
\label{sec:pheno}

\subsection{Background and model}

Here, to represent the situation in which a flagellum is attached to an organism's 
surface, we focus on the case of a finite beating filament, say $1$, pinned at its left end to a fixed support, with a free right end.  
As with all models for systems of this type, and the analysis in the previous 
section, we focus on low Reynolds number dynamics.
The structure of the most general equation of motion for a filament arising from balancing its tangential and normal forces
and bending moments is well known \cite{HinesBlum,BaylyWilson_bj,BaylyWilson_int,Machin_wave}.  Under the further assumption of linear filament
elasticity and resistive force theory, and assuming that the filament deviates only slightly from straight, the linearized 
equation of motion for the
tangent angle $\psi(s,t)$ as a function of the arc length $s$ and time $t$ takes the form
$\zeta_\perp \psi_t=af_{ss}-EI\psi_{ssss}$, where $E$ is the Young's modulus and $I$ the moment of inertia, per unit density, 
of the filament cross
section about the axis of rotation, and $f$ is the active bending moment.  Recognizing that within this approximation 
$\psi\simeq \partial_x h_1$ we obtain
\begin{equation}
\zeta_\perp \frac{\partial h_1}{\partial t}=a\frac{\partial f}{\partial x} -A\frac{\partial^4 h_1}{\partial x^4},
\label{general_eom}
\end{equation}
where $A=EI$ is the bending modulus of the filament.  The distinction between different  
models of active bending is to be found in the particular form of $f$, which may be coupled back to the geometry (e.g. $\psi$ and its
derivatives) through an auxiliary equation of motion.   

Recent work \cite{BaylyWilson_int} has studied the linearized dynamics of the unstable modes that arise in three models of
the form \eqref{general_eom}, known as sliding-control \cite{sliding_control}, curvature-control \cite{HinesBlum}, 
and the geometric clutch 
\cite{GC}.   The sliding control model,
whose equation of motion does not explicitly break left-right symmetry, was shown not to exhibit base-to-tip propagating modes of the 
kind seen in experiment. In contrast, both the curvature-control and geometric clutch models, which do display modes with the
qualitatively correct behavior, have a symmetry-breaking term  $\partial_{xxx}h_{1}$.   Furthermore, as the dynamics is translational invariant
in $y$, there can be no terms explicitly dependent on $h_1$ itself.
Motivated by these results, and with an eye toward the simplest PDE that will encode a characteristic wavelength, amplitude, and 
frequency of flagellar beating, we propose a form in which the left-right symmetry is broken with the derivative of the lowest
order possible, 
\begin{equation}
a\frac{\partial f}{\partial x}= -b\frac{\partial h_1}{\partial x}+G(\kappa),
\label{stuff}
\end{equation}
where $G$ is a nonlinear amplitude-stabilizing function of $\kappa\simeq \partial_{xx}h_{1}$, the filament curvature.  The presence of this advective
term does not reflect any explicit fluid flow, rather it encodes processes internal to the filament that break symmetry.
If we assume that the filament has $h_1\to -h_1$ symmetry, then $G$ will be
an odd function of its argument.  Expressing $G$ as a Taylor expansion up to cubic order we
arrive at the model of interest, henceforth called the advective flagella (AF) model
\begin{equation}
\zeta_{\perp} {\partial h_1 \over \partial t} = - C {\partial h_1 \over \partial x} - D {\partial^2 h_1 \over \partial x^2} -  A
{\partial^4 h_1 \over \partial x^4} + B
\left ( {\partial^2 h_1 \over \partial x^2}\right)^3~,
\label{onefilament00}
\end{equation}
where $A, B, C$ and $D$ are heuristic parameters of the model. We can now introduce dimensionless variables
$\xi = \alpha x$, $\tau = \varpi t$ and $ h = H h_1$, where $\alpha, \varpi$ and $H$ are constants. Direct 
substitution of these new variables into (\ref{onefilament00}), with the choice $c = C A^{1/2} (2/D)^{3/2}$, $\alpha = (D/2A)^{1/2}$, 
$\varpi = D^2/(4 \zeta_{\perp} A)$ and $H = A (2/BD)^{1/2}$ yields the characteristic length $\ell_c = \alpha^{-1} =
[A/(\varpi \zeta_{\perp})]^{1/4}$, and the dimensionless PDE
\begin{equation}
{\partial h \over \partial \tau} = - c {\partial h \over \partial \xi} - 2 {\partial^2 h \over \partial \xi^2} -  
{\partial^4 h \over \partial \xi^4} +
\left ( {\partial^2 h \over \partial \xi^2}\right)^3~,
\label{onefilament0}
\end{equation}
where $0\le \xi \le \Lambda$, with $\Lambda=\alpha L$.  The length $\ell_c$ is the well-known elastohydrodynamic penetration length
that arises in the study of actuated elastic filaments \cite{WG,Camalet}, and thus $\Lambda$ is the so-called Sperm number.
Of the many possible boundary conditions
we adopt the simplest: hinged at the left end ($h(0,\tau)=h_{\xi\xi}(0,\tau)=0$) and free at the right end ($h_{\xi\xi}(\Lambda,\tau)=
h_{\xi\xi\xi}(\Lambda,\tau)=0$).
In this rescaled form, the dynamics of a single filament is completely specified by $c$ and $\Lambda$.
Note that the dynamics (\ref{onefilament0}) does not enforce filament length conservation beyond linear order, but this should not 
introduce any qualitative changes in the results below. 

The linear operator in \eqref{onefilament0} is that of the advective version \cite{Burke} of the Swift-Hohenberg model \cite{SH}
without the term linear in $h$, and is identical to that in the phenomenological model of dendrite dynamics introduced by
Langer and M{\"u}ller-Krumbhaar \cite{LangerMK} and studied by Fabbiane, {\it et al.} in the context of control theory \cite{Fabbiane}.  
Heuristically, we recognize that the second and fourth derivatives naturally
select a most unstable length scale for a linear instability of the state $h=0$, the advective term leads to rightward wave propagation, 
and the nonlinearity leads to amplitude saturation. The intuitive understanding that these are the minimum necessary, but also sufficient, ingredients 
for a model of the beating of a single flagellum, is proven to be correct by the numerical and analytic work
presented in the following subsections. Moreover, in Section IV, we will also show that
the dual features of linear terms $-2h_{\xi\xi}-h_{\xi\xi\xi\xi}$ which lead to a band of unstable modes, and amplitude saturation through 
nonlinearity will allow for waveform compliance when two filaments are hydrodynamically coupled.  

\subsection{Numerical studies of a single filament}

In this section we present the dynamical behavior of the AF model obtained through numerical computations.  
The model \eqref{onefilament0} was solved numerically using an implicit finite-difference scheme described in detail by 
Tornberg and Shelley \cite{TornbergShelley}, including one-sided stencils for derivatives at the filament endpoints.
We shall see that for any choice of $c$ and $\Lambda$ the model selects a wavelength $\lambda$ (wavenumber $k=2\pi/\lambda$), frequency $\omega$ and maximum amplitude $A$.  Of 
particular interest are the limiting regimes $\Lambda/\lambda\sim 1$ and $\Lambda/\lambda \gg 1$.  The former is the regime seen in many
experiments (Fig. \ref{fig1}), while the latter corresponds to a semi-infinite system whose behavior far from the 
pinning wall approximates a traveling wave.

\begin{figure}[t]
\centering
\includegraphics[width=0.75\textwidth]{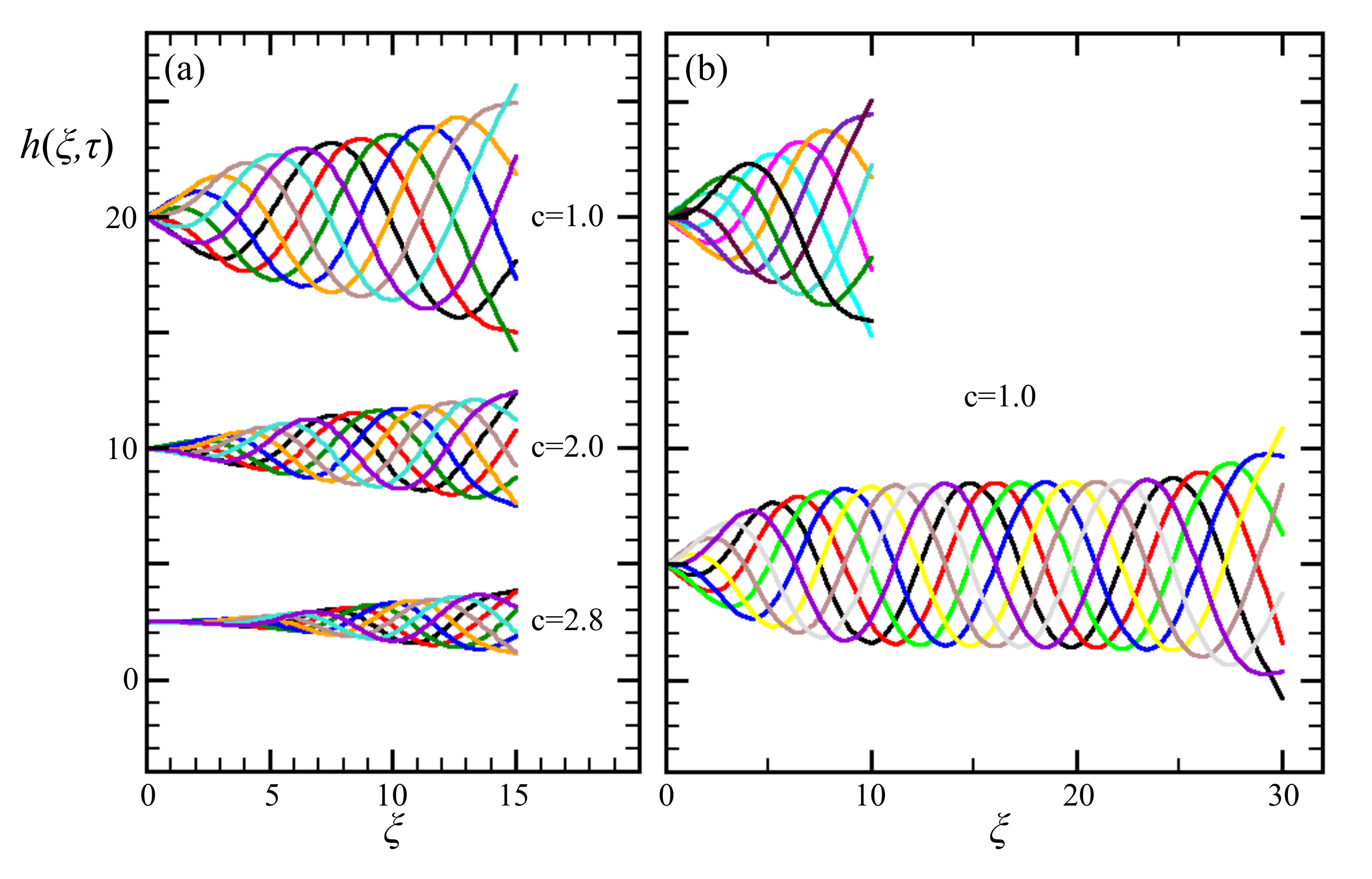}
\caption{Numerical results for the AF model of a single filament. (a) Overlaid waveforms during a single beating period 
for filament length $\Lambda=15$ and three values of $c$.  (b) As in (a) but for fixed $c=1$ and $\Lambda=10$ and $30$.}
\label{fig3}
\end{figure}

Following a short transient, we find that the filament evolves toward a periodic waveform of nonuniform, finite amplitude.
Figure \ref{fig3}a shows overlays of $h(\xi,\tau)$ at various times during a full cycle for $\Lambda=15$ and three values of $c$.
We observe that for this length, and $c=1$ (top waveform) the oscillation amplitude has not reached saturation at the free end.
Thus, even larger values of $c$ (middle and bottom) naturally 
advect the waveform even faster to the right, resulting in a smaller maximum amplitude 
at the free end.  As seen by comparing the two examples in Fig. \ref{fig3}b with the top in Fig. \ref{fig3}a, where all three filaments
have $c=1$ but different length, for $\Lambda=10$ the amplitude envelope is clearly linear in $\xi$, for $S=15$ there is the beginning of a rollover,
and for $\Lambda=30$ there is clear saturation.
For $\Lambda/\lambda \sim 1$ the maximum amplitude of the waveform is reached at the free end of the filament, as seen
in the natural biological waveforms and also in the models described above \cite{BaylyWilson_int}. 
In general, for a given value of
$c$, as $\Lambda$ grows relative to $\lambda$, the filament displays two distinct regions: a transition region adjacent to the wall where the 
amplitude grows along $\xi$ up to a characteristic ${\cal L}$, and a region beyond where the amplitude saturates and the oscillations 
approximate a traveling wave. The `healing length' ${\cal L}$ increases with $c$.

In the regime $\Lambda/\lambda\gg 1$, and far from the wall, the amplitude $A$, wave vector $k$, and frequency $\omega$ of the 
approximate traveling wave are determined by $c$ only.  The solid symbols in Figure \ref{fig4} indicate those numerical 
results.  To put these results in context, note that a linear stability analysis of the operator $-2h_{\xi\xi}-h_{4\xi}$ 
leads to a growth rate which is maximized at $k^*=1$ where, without loss of generality, we consider only the positive branch of
solutions. We see from Fig. \ref{fig4}a that the value of $k$ selected by the system is always less than $1$, but $k\to 1$ as $c$ increases,
and also, from Fig. \ref{fig4}b, that the selected frequency is consistent with the relation $\omega\simeq ck$.  Finally, the saturated amplitude 
$A$ of the waveform far from the wall is a strongly decreasing function of $k$ (Fig. \ref{fig4}c).

\subsection{Approximate analytical solution}

\subsubsection{Amplitude saturation} 
\begin{figure}[t]
\centering
\includegraphics[width=1.0\textwidth]{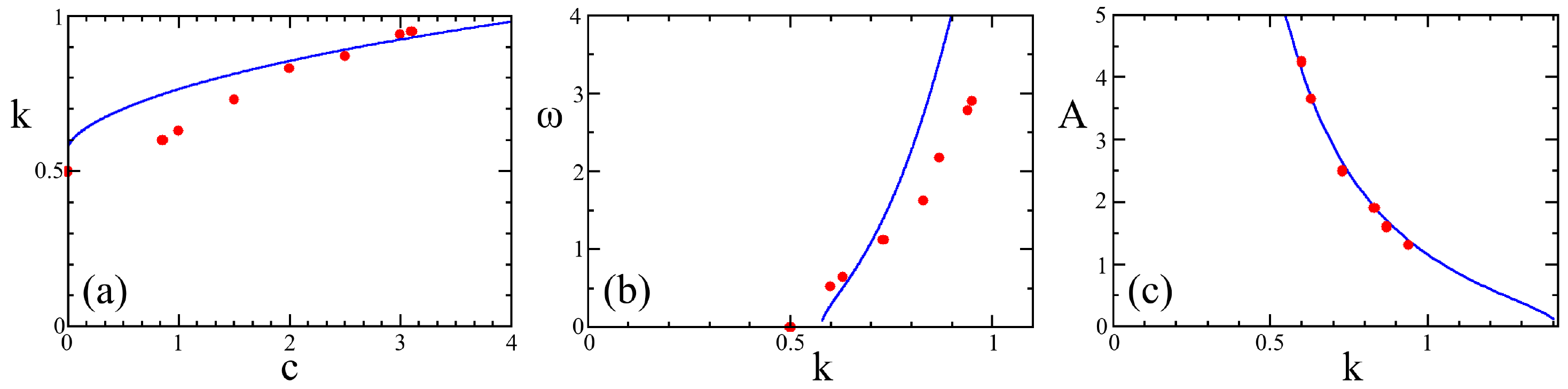}
\caption{Waveform selection in the AF model. Numerical (red circles) and analytical (blue lines) results for long filaments:
(a) selected wave vector $k$ vs. advective parameter $c$, (b) selected frequency $\omega$ vs. $k$, and (c) saturated amplitude $A$ vs. $k$.}
\label{fig4}
\end{figure}

Far away from the wall the solution of \eqref{onefilament0} is well approximated by 
a traveling wave with a time dependent amplitude of the form
\begin{equation}
 h(\xi,\tau)= A(\tau) \cos (k\xi -\omega \tau)~.
\label{travellingsolution}
\end{equation}
Substituting (\ref{travellingsolution}) into (\ref{onefilament0}), 
rewriting the cubic cosine term with the multiple angle formula, ignoring terms involving ${\cal O} [3(k\xi -\omega \tau)]$, and 
equating the coefficients of the sine and cosine to zero
yields an evolution equation for $A(\tau)$ and a relationship between the 
wavenumber $k$ and the angular frequency $\omega$, namely
\begin{subeqnarray}
\label{ftequation}
\omega &=& c k, \\
A_{\tau} &\simeq& k^2 (2 - k^2)A - {3 \over 4} k^6 A^3~.
\end{subeqnarray}
The evolution equation for the amplitude $A(\tau)$ is of the Bernoulli type and can be easily solved after
making the change of variables $z = A^{-2}$, 
with solution of the form $z = B {\rm e}^{\beta \tau} + \gamma $, where $B$, $\beta$, and $\gamma$ are 
constants, corresponding to 
a time dependent amplitude given by
\begin{equation}
A(\tau) = \left (B {\rm e}^{- 2 ( 2 k^2 - k^4) \tau}  + {3 k^4 \over 4(2 -k^2)} \right)^{-1/2}.
\label{solutionforf}
\end{equation}
Provided $0 < k < \sqrt 2 $, $A$ has a finite value as $\tau \to \infty$, namely
\begin{equation}
A_{\infty}(k) = {2 \sqrt{2 - k^2} \over \sqrt 3 k^2 }.
\label{maxamplitude}
\end{equation}
$A_{\infty}(k)$ is shown as the solid line in Fig. \ref{fig4}c, and it is in excellent agreement with the numerical data for $A$ expressed 
now as a function of the numerically selected $k$.  We note that the steady-state solution can be found exactly
in terms of elliptic integrals by transforming into a moving coordinate system with $X=\xi-c\tau$, $h(\xi,\tau)\to F(X)$ and solving 
$-2G-G_{XX}+G^3=0$
for $G=F_{XX}$.  However, the slight increase in accuracy that this approach produces comes at the expense of a lack of clarity compared
to the one-mode approximation \eqref{travellingsolution}.

\subsubsection{Wavenumber and frequency selection} 

While the traveling wave solution is a good approximation far away from 
the wall, it only provides an asymptotic relationship between the wavenumber and the frequency 
but does not yield any information about the possible numerical values of $k$ and $\omega$. 
In order to find these numerical values the linearized evolution equation 
\begin{equation} 
{\partial h \over \partial \tau} = - c {\partial h \over \partial \xi} - 2 
{\partial^2 h \over \partial \xi^2} -  {\partial^4 h \over \partial \xi^4},
\label{linear}
\end{equation}
may be used. Besides the trivial solution $h(\xi,\tau) = h_0$ where $ h_0$ is a constant, (\ref{linear}) has a solution of 
the form $h(\xi,\tau) = {\rm exp}(\Omega \tau +K \xi) $, where $\Omega = \sigma +i \omega^{\prime}$ 
and $ K= p+ik $.
In the present case, the only solutions of interest are those with a positive temporal growth rate $\sigma > 0$ because any other solution 
will decay to $h(\xi,\tau) = h_0$. Moreover, we expect the solution corresponding to the maximum growth 
rate to be the one that will control the evolution of the shape $h(\xi,\tau)$. Hence, to find the extremal values of $K$ and 
$\Omega$ we replace $h(\xi,\tau)$ into (\ref{linear}) and then, from the characteristic equation, calculate the derivative of $\Omega$ 
respect of $K$:
\begin{equation}
\Omega = -c K -2 K^2 -K^4 \ \ {\rm and} \ \  {d\Omega \over dK} = -c - 4K -4K^3 .
\end{equation}
After separating real and imaginary parts they become:
\begin{subeqnarray}
\label{maxK}
\sigma &=& -cp - 2(p^2 - k^2) - [(p^2 -k^2)^2 - 4 p^2 k^2], \\
\omega^{\prime} &=& -c q -4 p k - 4p^3 k + 4 p k^3, \\
0 &=& -c - 4p -4p^3 +12 p k^2  \ \ \ {\rm and} \\
{\partial \omega^{\prime} \over \partial p} &=& -4 k - 12 p^2 k - 4 k^3, 
\end{subeqnarray}
where we have already set the real part of the derivative of $\Omega$ equal to zero and note
that the complex part of the derivative is identical to calculating ${\partial \omega^{\prime}/\partial p} $.
In general, $ \partial \omega^{\prime} / \partial p $ is also set to zero to find the boundary between 
convective and absolute instability \cite{Tobias,Saarloos}. Here this second constraint has been relaxed so that it is possible to find
the curve $ \omega^{\prime}(k)$ along which the system is most unstable. In fact, rewriting
equation (\ref{maxK}c) as $-12 pk^2 = -c -4p - 4 p^3 $ and replacing
this expression into (\ref{maxK}b) yields $\partial\omega^{\prime}/\partial p = - 8k^3$.  Equating
the two expressions for $ {\partial \omega^{\prime} /\partial p}$ then gives
\begin{equation}
p^2 = k^2 - {1 \over 3}.
\label{pofk}
\end{equation}
Replacing (\ref{pofk}) back into equations (\ref{maxK}c-a) gives 
the expressions for the velocity $c$, the angular frequency $\omega = - \omega^{\prime}$  
and the growth rate $\sigma$ as functions of the wave number $k$:
\begin{subeqnarray}
c &=&  8 \left( k^2 - {1 \over 3} \right)^{3/2}, \\
\omega &=& 8 k^3  \left( k^2 - {1 \over 3} \right)^{1/2},\\
\sigma &=& -4 k^4 + 4 k^2  - {1 \over 3}.  
\end{subeqnarray}
From these results it becomes clear that the only solutions which satisfy the hypothesis 
for $h(x,t)$ are those with values $k \ge 1/{\sqrt 3}$ because for values of $k < 1/{\sqrt 3}$
both, $c$ and $\omega$ are complex. In particular, the value $ k = {1 /{\sqrt 3}}$ corresponds to 
$ c = \omega = 0$.  As seen in Fig. \ref{fig4}b the numerical counterpart
to this boundary is $ k = 1/2 $. While this represents a discrepancy of approximately 15 percent 
between the analytic prediction and numerical results, it can be seen in the figure that the basic trend in the selected frequency is
captured.  Figure \ref{fig4}a shows the comparison between the numerical $k(c)$ and the analytical result above, and in this case
larger the value of $c$ the closer 
the two become.  Because the analytic expression (\ref{maxamplitude}) for the maximum amplitude as a function of $k$ 
was obtained using the full nonlinear equation, this prediction shows almost perfect agreement with the 
numerical results (Fig. \ref{fig4}c). 

Another value of interest is the critical $k^*$ with zero growth rate, and  which can be found by solving 
the bi-quadratic equation $4 k^4 - 4 k^2  + {1 \over 3} = 0 $ which has 
only one real root above the threshold for $k$:
\begin{equation}
 k^* = \left[ {1 + {\sqrt {(2/3)}}\over 2} \right]^{1/2} \simeq 0.953
\end{equation}
All solutions in the interval $ 1/{\sqrt 3} \le k \le k^* $ have positive growth rates 
with values $ \sigma(k^*)  \le \sigma \le  \sigma (1/{\sqrt 2}) $ where 
$ \sigma(k^*) = 0$ and  $\sigma (1/{\sqrt 2}) = 2/3 $;
while  all solutions for which $k > k^*$ have a negative growth rate and, hence, relax back to the
solution $h = h_0 $. 
The critical angular frequency and velocity  that 
correspond to $\sigma (k^*) = 0$ are
\begin{subeqnarray}
\label{criticalomegaandc}
\omega (k^*) &=&  {2 \over  {\sqrt 3}} \left( 1 +  {\sqrt { 2 \over 3}} \right)^{3/2} 
\left( 1 + 3 {\sqrt { 2 \over 3}} \right)^{1/2}  \simeq 5.250, \\
c(k^*) &=&  {4 \over 3 {\sqrt 6}}\left( 1 + 3 {\sqrt { 2 \over 3}} \right)^{3/2}  \simeq 3.487.
\end{subeqnarray}
The numerical value of the velocity for which the oscillatory solutions ``disappear"  is $c^* \simeq 3.2$, and
the analytic prediction (\ref{criticalomegaandc}b) is well within 
the acceptable limits of agreement for the approximation.  Moreover, the analytic prediction for 
the functional form of the ratio $\omega(k)/ c(k)$ is qualitatively correct and given by
\begin{equation}
{\omega \over c} (k) = {k^3 \over k^2 -{1\over3}},
\end{equation}
which as expected, tends to $ \omega/c  \to k $ as $k$ grows.

\subsubsection{Frequency-amplitude relation} 

\begin{figure}[t]
\centering
\includegraphics[width=0.95\textwidth]{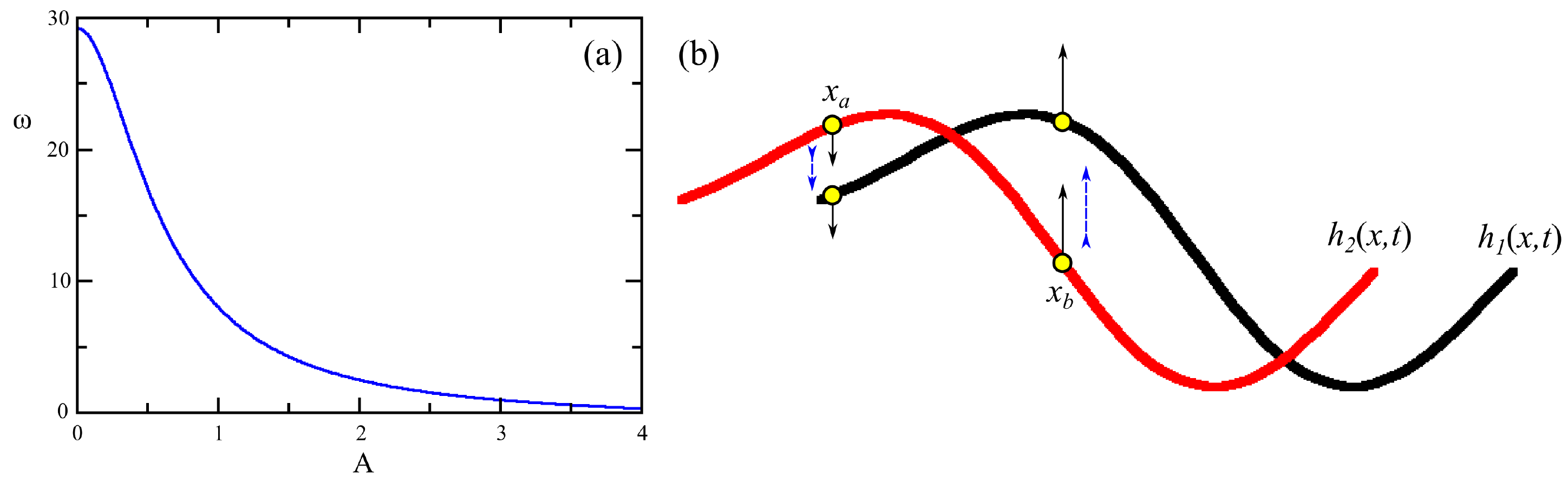}
\caption{Mechanism of synchronization. (a) Amplitude dependence of traveling-wave frequency $\omega$.  (b) Schematic of two 
nearby filaments, with black arrows indicating the direction of motion of points on the filaments, and blue arrows showing
the direction of fluid motion induced by filament $1$.}
\label{fig5}
\end{figure}

Before proceeding to numerical studies of coupled filaments we are in a position to see how synchronization of two nearby filaments
may occur.  First, we note that when two filaments beat in synchrony the fluid gap between them is nearly constant, whereas when they 
are out of synchrony
the gap varies with position. That variation produces fluid forces that will deform the filaments such that their local amplitude and frequency 
will be altered.  As seen in bead-spring models \cite{NEL}, for example, 
a stable synchronous state may occur when the frequency is a decreasing function of 
amplitude.  In the case of the approximate traveling-wave states we have
discussed in this section, the results above can be used directly to calculate $\omega(A)$, as shown in Fig. \ref{fig5}a,
\begin{equation}
\omega(A)={32\over 9A^4}\left[\sqrt{1+6A^2}-\left(1+{A^2\over 2}\right)\right]^{1/2}\left[\sqrt{1+6A^2}-1\right]^{3/2}.
\label{omegavsamp}
\end{equation}
To see how the synchronization mechanism operates within the present model, and anticipating the numerical results presented below, 
consider the configuration of two filaments shown in Fig. \ref{fig5}b, each traveling to the right,
where the black arrows indicate the local direction of motion of the filaments at two distinct coordinates, and the blue arrows indicate the
direction of the fluid flow induced by filament $1$ ($h_1(x,t)$) on $2$.
At the point $x_a$ the fluid flow acting on filament $2$ will push it further down, whereas at point $x_b$ that flow will pull it
upwards.  The net effect is that the local wave amplitude of filament $2$ will be decreased, and by the relationship in Fig. \ref{fig5}a
its frequency will increase, moving it faster to the right and hence catching up with filament $1$.  Similar considerations show that
the effect of filament $2$ on $1$ is to increase its amplitude, hence to decrease its frequency.  Thus, the stable state is the 
in-phase synchronized one.  This elastohydrodynamic mechanism is the continuum analog of that which operates in bead-spring models.

\section{Two coupled filaments}

\subsection{Symmetry and stability considerations}

In this section we discuss the coupled dynamics of two nearby filaments using the model of Section \ref{sec:pheno}
with the coupling derived in Section \ref{sec:asym}.  When two coupled filaments have the same intrinsic speeds 
$c_1= c_2 = c $ then they obey
\begin{subeqnarray}
\label{coupled}
{\partial {\hat h}_1 \over \partial \tau} &=& {\cal N}_c ({\hat h}_1) + \epsilon  {\partial {\hat h}_2 \over \partial \tau}, \\
{\partial {\hat h}_2 \over \partial \tau} &=& {\cal N}_c ({\hat h}_2) + \epsilon  {\partial {\hat h}_1 \over \partial \tau}, 
\end{subeqnarray}
where, following Eq. \ref{onefilament0},  the nonlinear operator ${\cal N}_c ({\hat h}_i)$ is 
\begin{equation} 
{\cal N}_c ({\hat h}_i) = - c {\partial {\hat h}_i \over \partial \xi} - 2 
{\partial^2 {\hat h}_i \over \partial \xi^2} -  {\partial^4 {\hat h}_i \over \partial \xi^4} +
\left ( {\partial^2 {\hat h}_i \over \partial \xi^2}\right)^3.
\label{operator}
\end{equation}

By direct substitution, it is straightforward to show that Eqs. \ref{coupled}, with the operator \ref{operator}, have two exact solutions
$S_1 : ( {\hat h}_1(\xi,\tau), {\hat d} + {\hat h}_1(\xi,\tau))$ (sinuous solution), and 
$S_2 : ( {\hat h}_1(\xi,\tau), {\hat d}- {\hat h}_1(\xi, \tau))$ (varicose solution), 
where ${\hat d}$ is a dimensionless constant, provided ${\hat h}_1(\xi,\tau)$ satisfies the nonlinear, autonomous equation
\begin{equation} 
(1 \mp \epsilon) {\partial {\hat h}_1 \over \partial \tau} 
= - c {\partial {\hat h}_1 \over \partial \xi} - 2 
{\partial^2 {\hat h}_1 \over \partial \xi^2} -  {\partial^4 {\hat h}_1 \over \partial \xi^4} +
\left ( {\partial^2 {\hat h}_1 \over \partial \xi^2}\right)^3,
\label{h1}
\end{equation}
which is the rescaled PDE that governs a single isolated filament. 
Note that here the plus (minus) sign corresponds to the sinuous (varicose) configuration.
The effective times $\tau'=\tau/(1\mp\epsilon)$ correspond 
to faster motion than an isolated filament in the sinuous case and slower motion in the varicose 
arrangement.  This can be understood as a result of decreased and increased viscous dissipation in the two
cases, respectively, consistent with the results of the waving-sheet model.

To study the stability of these solutions a small perturbation is introduced.
Because our focus is on the difference between the filament positions, it is sufficient to examine perturbed solutions 
of the form $\left( {\hat h}_1(\xi,\tau), {\hat d} \pm {\hat h}_1(\xi, \tau) + \eta(\xi,\tau)\right)$ where 
$\vert \eta(\xi,\tau)/{\hat h}_1(\xi,\tau) \vert \ll 1$.
Then, the linearized equation of motion for $\eta$ is
\begin{equation} 
(1 \pm \epsilon) {\partial \eta \over \partial \tau} = - c {\partial \eta \over \partial \xi} 
+ \left[\pm 3  \left( {\partial^2 {\hat h}_1 \over \partial \xi^2}\right)^2  -2 \right] {\partial^2 \eta \over \partial \xi^2} 
-  {\partial^4 \eta \over \partial \xi^4}.
\label{heq}
\end{equation}
Note that here the upper (lower) sign corresponds to the sinuous (varicose) configuration.
The dynamics (\ref{heq}) is close in form to the linearized dynamics of a single filament, with one crucial difference: and 
in the original operator the coefficient of the second spatial derivative was negative
(namely $-2$) which corresponds to the ``anti"-diffusion that produces the instability, whereas now, the 
base solution parametrically forces the perturbation through the coefficient of the second derivative.
Thus, depending on the characteristics of the 
solution ${\hat h}_1$, and only in the sinuous case, it is possible to find regions with a positive effective diffusion coefficient.

As pointed out in the previous section, the asymptotic solution in time of (\ref{h1}) far away from the origin 
is a traveling wave ${\hat h}_1(\xi,\tau) \simeq A_{\infty} \cos (k\xi -\omega \tau) $ with $A_{\infty} = 2(2 - k^2)^{1/2} / k^2$,
which yields
\begin{equation} 
\left< \pm 3  \left( {\partial^2 {\hat h}_1 \over \partial \xi^2}\right)^2  - 2 \right>_{\xi,\tau} = \pm 2\left( 1 - k^2 \right)
\label{coeffavge}
\end{equation}
as the average coefficient of the diffusive term in Eq. \ref{heq}.  This coefficient is positive (hence
stabilizing) for the sinuous configuration 
and negative (destabilizing) for the varicose one.  In the presence of the always stabilizing influence of the
fourth derivative, this implies that the only linearly stable configuration is the sinuous one.

When the speeds of the filaments are slightly different,
$c_1 = c - (\Delta c / 2)$ and $c_2 = c + (\Delta c / 2)$  with $\Delta c / c \ll 1$, 
a similar analysis to the one described above makes it possible to find an approximate solution that corresponds to
a quasi-sinuous configuration, and for which the first order solution is the same as when $\Delta c=0$.  
The initial system that governs the motion is
\begin{subeqnarray}
\label{differentc}
{\partial {\hat h}_1 \over \partial \tau} &=& {\cal N}_{c_1} ({\hat h}_1) + \epsilon  {\partial {\hat h}_2 \over \partial \tau}, \\
{\partial {\hat h}_2 \over \partial \tau} &=& {\cal N}_{c_2} ({\hat h}_2) + \epsilon  {\partial {\hat h}_1 \over \partial \tau}, 
\end{subeqnarray}
and the proposed solution is $ S: ({\hat h}_1 = {\bar h} -\eta , {\hat h}_2 = {\bar h} +\eta)$ where $ \eta/{\bar h} \ll 1$. Direct substitution and some simple
algebra yield the evolution equation for ${\bar h}$,
\begin{equation}
( 1 - \epsilon) {\partial {\bar h} \over \partial \tau} ={\cal N}_c ({\bar h}) + 
 3  \left({\partial^2 \eta \over \partial \xi^2}\right)^2 
{\partial^2 {\bar h} \over \partial \xi^2} - \left({\Delta c \over 2}\right) {\partial h \over \partial \xi}.
\label{differentchbar}
\end{equation}
Keeping only linear terms in $\eta $ this reduces to the original autonomous one for a single filament, but with 
coefficient $1 - \epsilon $ in front of the temporal derivative.  The associated linearized equation for $\eta$ is
similar to (\ref{heq}), but also has a forcing
\begin{equation} 
(1 + \epsilon) {\partial \eta \over \partial \tau} = - c {\partial \eta \over \partial \xi} 
+ \left[3  \left({\partial^2 {\bar h} \over \partial \xi^2}\right)^2
- 2 \right] {\partial^2 \eta \over \partial \xi^2} -  {\partial^4 \eta \over \partial \xi^4}
-\left({\Delta c \over 2}\right) {\partial {\bar h} \over \partial \xi}
\label{heqdifferentc}
\end{equation}
Replacing into (\ref{heqdifferentc}) the traveling wave solution (\ref{travellingsolution}) we obtain
\begin{equation} 
(1 + \epsilon) {\partial \eta \over \partial \tau} + c {\partial \eta \over \partial \xi} 
- \left[ 4 (2 -k^2) \cos^2(k\xi -\omega \tau)
- 2 \right] {\partial^2 \eta \over \partial \xi^2} -  {\partial^4 \eta \over \partial \xi^4}
\simeq  \Delta c  { \sqrt{ 2 -k^2} \over \sqrt 3 k^3} \sin (k\xi -\omega \tau)
\label{driveneqn}
\end{equation}
Notice that the diffusion coefficient and the forcing are out of phase. When $\vert \cos ( k\xi -\omega \tau)\vert $ is large, the diffusion
coefficient is positive and $\eta $ decreases making ${\hat h}_1 \simeq {\hat h}_2$; these are the regions where the forcing is very small and 
tends to leave the system unperturbed. In the regions of space and time where the cosine is very small and the diffusion 
coefficient takes a negative value, the forcing is strong and in opposition to the growth of $\eta $, and even though the filaments can 
never be in absolute synchrony, as when the two velocities are the same, the forcing keeps the asynchrony to a minimum.

\subsection{Numerical results}

\begin{figure}[t]
\centering
\includegraphics[width=0.95\textwidth]{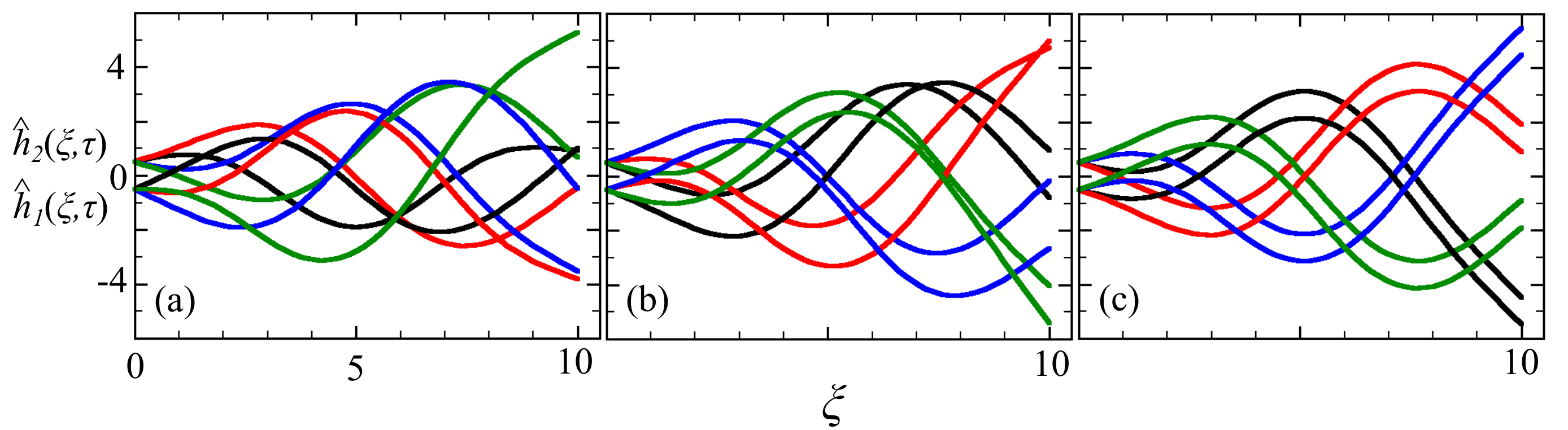}
\caption{Numerical synchronization dynamics for two filaments with $\Lambda=10$ and $c=1$.  Panels (a)-(c) show overlaid waveforms
at four points within a single oscillation cycle at early (a), middle (b) and late (c) times.}
\label{fig6}
\end{figure}

Numerical studies of the coupled dynamics \eqref{coupled} of two filaments with the same speed parameter $c$ show robust synchronization
for a broad range of initial conditions.  Figure \ref{fig6} shows the evolution toward synchrony of a pair of filaments with $c=1$, $\Lambda=10$, 
and $\epsilon=0.1$, computed over a total time of $T=60$.  Panel (a-c) show
the two filaments at early ($2\le \tau \le 10.7$), intermediate ($16.3\le \tau \le 25.6$), and late ($48.7\le \tau \le 57.3$) times and at 
four equally spaced time intervals within each corresponding cycle.  From the clearly disordered pattern in (a) the filaments evolve
to a fully-synchronized state in (c).  

\begin{figure}[t]
\centering
\includegraphics[width=0.6\textwidth]{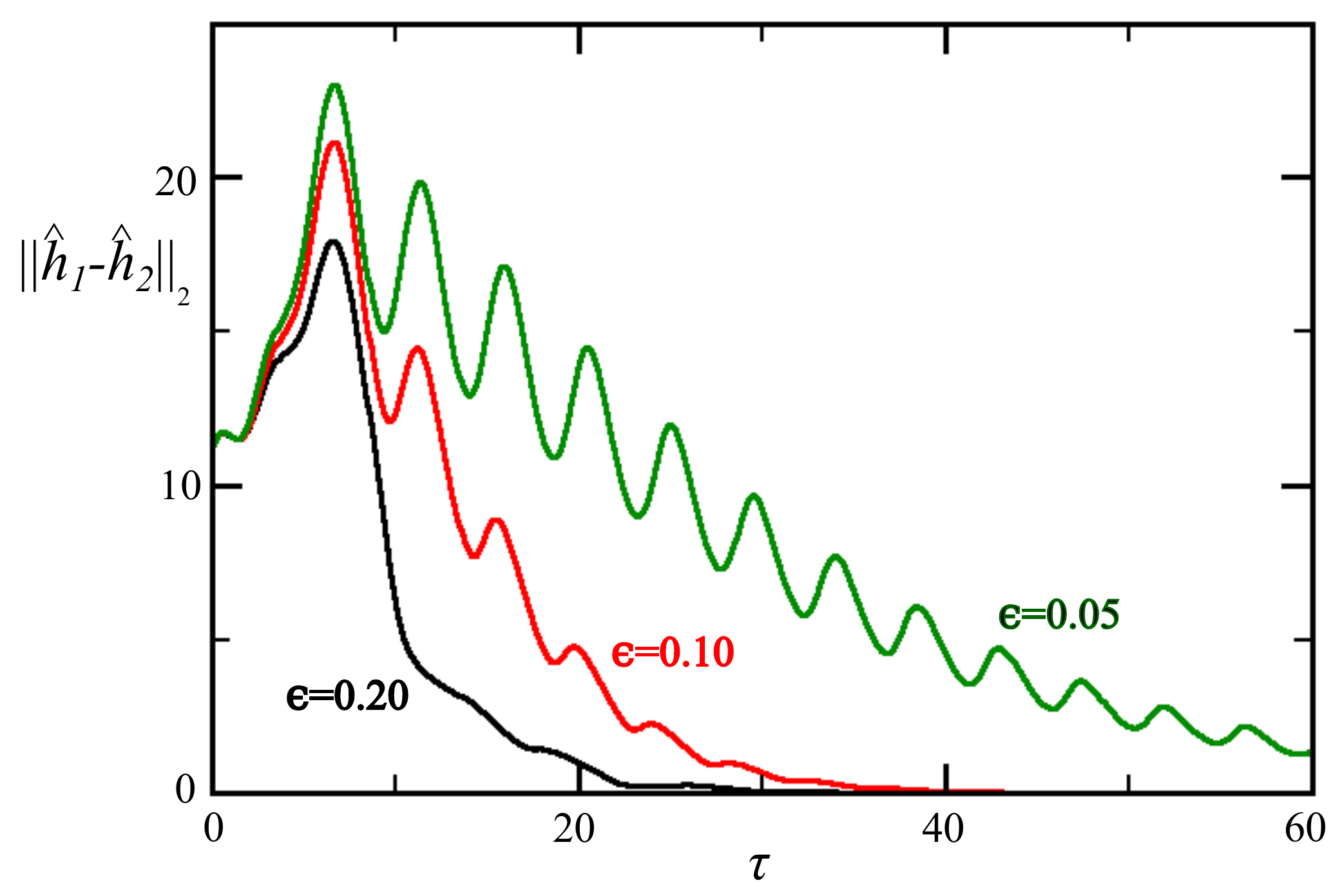}
\caption{Approach to synchrony.  The $L_2$ norm of the difference in waveform between two filaments as a function of time, 
for $\Lambda=10$, $c=1$, and indicated values of $\epsilon$.}
\label{fig7}
\end{figure}

A convenient quantity to characterize the degree of synchrony of two filaments is the $L_2$ norm of the difference 
of their displacements, $D =\parallel\!{\hat h}_1(\xi, \tau)-{\hat h}_2(\xi, \tau)\!\parallel_2$.  
Figure \ref{fig7} shows the temporal behavior 
of $D$ displacements, 
for three different values of the coupling constant $\epsilon$.  While the rate
of synchronization increases with $\epsilon$, and the details of the decay of the norm differ, synchrony occurs in all cases.
Further analysis shows that, when averaged over the fast oscillation, the approach to synchrony is exponential, as would be expected from 
the fact that the 
linearized dynamics close to synchrony is first order in time.
In the physical regime, with $\epsilon \sim 0.2$-$0.5$, we see in Fig. \ref{fig7} that synchrony occurs in a 
matter of a few beat cycles.  This rapid synchronization is often seen in biological systems, including {\it Chlamydomonas} flagella
subjected to hydrodynamic perturbations \cite{WanG} and during the photoshock response \cite{unpub}.

\begin{figure}[h]
\centering
\includegraphics[width=0.6\textwidth]{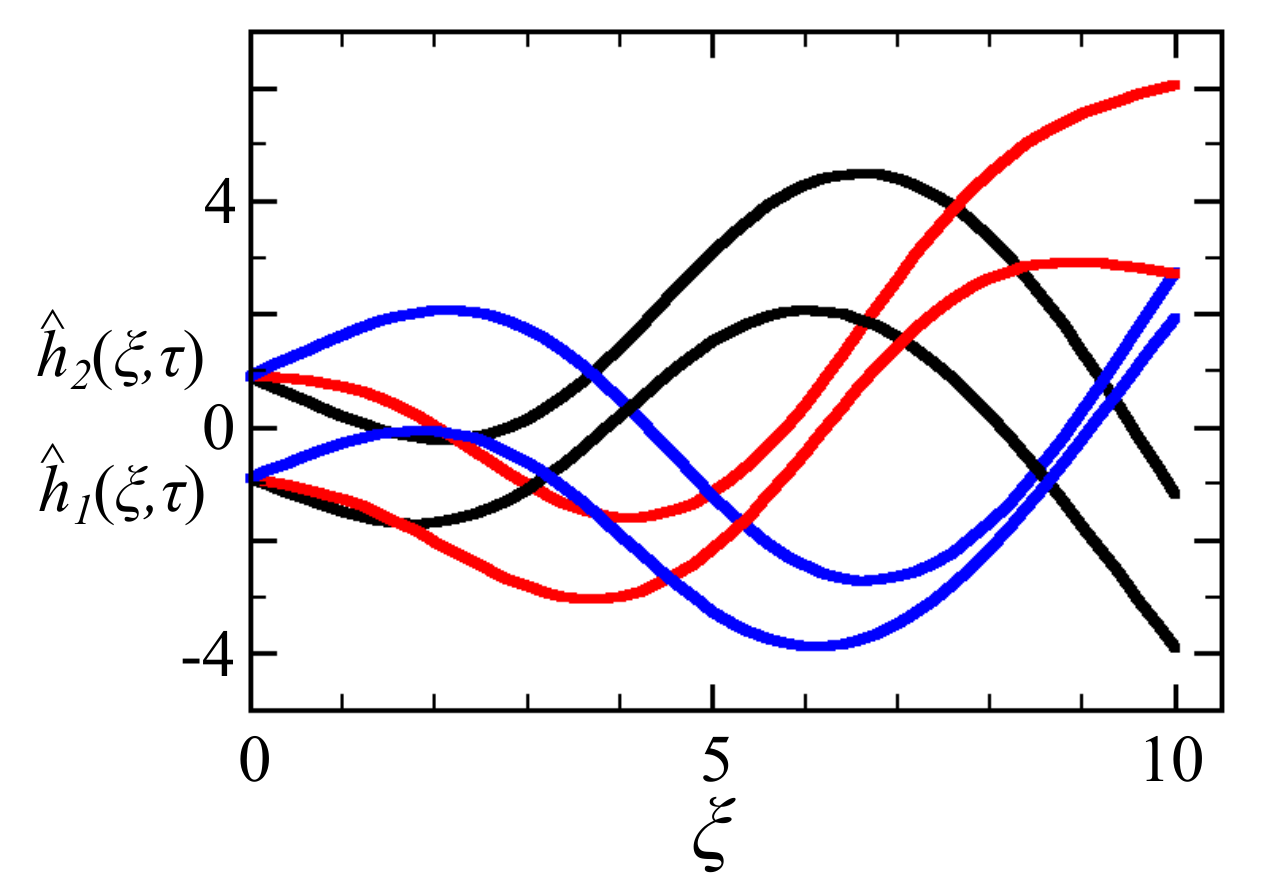}
\caption{Synchronization with unequal advective coefficients.  Overlaid waveforms when frequency locking has 
been achieved, for $c_1=0.9$ and $c_2=1.0$ and $\Lambda=10$.  The faster flagellum (upper) is phase-shifted forward with respect to the 
slower one.}
\label{fig8}
\end{figure}

Finally, we discuss the case with slightly different values of the speed parameter $c$.  We expect that the two 
filaments will frequency lock but display a finite phase shift.  This is borne out in the numerical studies, as shown in Fig. \ref{fig8}
for $c_1=0.9$ and $c_2=1.0$, where the upper filament ($2$) leads the lower one ($1$), while the two display identical frequencies.

\section{Discussion}
\label{sec:conc}

In this work we have presented two main results.  First, under the assumptions of coplanarity and small-amplitude deformations,
we have derived the leading order coupling term that describes the 
hydrodynamic interaction between two nearby slender bodies whose separation $d$ lies within the asymptotic limit $a\ll d\ll L$.  
This is a very general result formally
expressed simply in terms of the velocities of each filament, whatever their microscopic origins.  
Second, we have applied this result to a model of flagellar beating that has the minimum necessary features to capture the
essence of the system: self-sustained oscillations, broken left-right symmetry, bending elasticity, and waveform 
amplitude saturation.  Analytical and numerical studies of the model for the case of single filaments 
illustrate the mechanisms of wavelength, frequency, and amplitude selection, while those for coupled pairs display the basic
elastohydrodynamic mechanism of synchrony.

We emphasize that because of the extremely general form of the inter-filament coupling term, it can immediately 
be used to extend any particular model of the beating of a single filament to the coupled dynamics of two or more.
A worthwhile extension would be the generalization of this result to include the presence of a no-slip wall at which filaments are
anchored, with the goal of understanding metachronal waves.  
Finally, we expect the approach to flagellar dynamics outlined here, based on long-wavelength expansions and symmetry considerations, 
and reduced to a single autonomous PDE, will prove useful in the analysis of experimental waveforms and dynamics.

\section{acknowledgments} We are grateful to Andreas Hilfinger and Wim van Saarloos for very useful conversations.  This work was supported by
a Wellcome Trust Senior Investigator Award (REG \& AIP) and by a Marie Curie Career Integration Grant (EL).

\bibliographystyle{aipauth4-1}

\end{document}